\documentclass[11pt,twoside]{article}

\usepackage{asp2006}
\usepackage{graphicx}
\usepackage{epsfig}
\usepackage{lscape}

\newcommand{\mo}{\ifmmode ^{-1}\else $^{-1}$\fi}
\newcommand{\mt}{\ifmmode ^{-2}\else $^{-2}$\fi}
\newcommand{\microns}{\ifmmode \,\mu$m$\else \,$\mu$m\fi}
\newcommand{\WHzsr}{\ifmmode $\,W\,Hz\mo\,sr\mo$\else \,W\,Hz\mo\,sr\mo\fi}

\begin{document}
\title{{\it Spitzer} Observations of Powerful Radio Sources}   
\author{\author{K.\ Cleary$^1$, C.\ R.\ Lawrence$^1$, J.\ A.\ Marshall$^2$, L.\ Hao$^2$ and D.\ Meier$^1$}}   
\affil{$^1$ Jet Propulsion Laboratory, California Institute of
Technology \\
$^2$ Astronomy
department, Cornell University, Ithaca, NY 14853}

\begin{abstract} 
We have measured the mid-infrared radiation from an orientation-unbiased sample of 3C RR galaxies and quasars with $0.4<z<1.2$ using the IRS and MIPS instruments aboard the {\it Spitzer Space Telescope}.  We fit the {\sl Spitzer\/} data as well as
other measurements from the literature with synchrotron and dust components. At 15\microns, quasars
are typically four times brighter than radio galaxies with the same isotropic
radio power.  Based on our fits, half of this difference can be attributed to the presence
of non-thermal emission in the quasars but not the radio galaxies.  The other half
is consistent with dust absorption in the radio galaxies but not the quasars.
\end{abstract}

\section{Introduction}

Only a small fraction of AGN produces the prodigious radio power of FR\,II galaxies
and quasars.   Nevertheless, the fact that FR\,IIs can be
studied in an orientation-unbiased sample, selected on the basis of low-frequency radio emission, makes them uniquely valuable in separating the
effects of orientation from physical differences. The mid- and far-infrared properties of these powerful radio sources are largely
unknown.  Their space density is so low that only a few (e.g., 3C \,405 = Cygnus
A) are at low redshifts. The {\it Spitzer Space Telescope} \citep{werner_etal_04} promised a major advance
in sensitivity over previous telescopes, and the capability to measure these
objects. We therefore undertook observations with {\em Spitzer\/} of powerful radio
sources.  The overall goal was straightforward: to measure for the first time the mid-
and far-infrared emission from an orientation-unbiased sample of the powerful
radio sources.

This paper gives a brief overview of the observations, data reduction and some of the main results. Further details may be found in \cite{cleary_etal}.

\section{The Sample\label{sample}}

We require a sample selected at low frequency, with  $L_{\rm 178\,MHz} > 10^{26}$\,\WHzsr\ and with a reasonable balance between FR\,II radio galaxies and quasars.

The sample of \cite{barthel_89} provides the ideal starting point.  It contains the 50
sources in the complete low-frequency 3C RR catalog of \cite{laing_etal_83} with
$0.5 \leq z \leq 1.0$.  All have emitted radio luminosity $L_{\rm 178\,MHz} > 10^{26}$\,\WHzsr. To reduce the observing time required, we reduced the sample to
33 objects based on ecliptic latitude.  We also added one source to the sample, 3C \,200, at
$z=0.458$, because we already had a 16\,ks {\it Chandra\/} observation of it.  The {\em Spitzer\/} sample consists of 16~quasars and 18~galaxies.

\section{Observations and Data Reduction\label{obs}}

All sources in the sample were scheduled for observation with the Long-Low module of the {\it Spitzer} Infrared Spectrograph \citep[IRS;][]{houck_etal_04} resulting in spectra in the range 15--37\microns , as well as with the Multiband Imaging Photometer \citep[MIPS;][]{rieke_etal_04} in photometry mode at 24, 70, and 160\microns . All objects in the sample are unresolved by the {\em Spitzer\/} instruments.

The IRS spectra were extracted from the basic calibrated data (BCD) images provided by the Spitzer Science Center, using
the Spectroscopy Modeling and Analysis Tool \citep[\textsc{Smart};][]{higdon_etal_04}. Mosaics of the MIPS 24 and 70\microns\ BCD images were produced using the Mosaicking and Point
Source Extraction tool, \textsc{Mopex} \citep{makovoz_khan_05}, while for 160\microns , the
post-BCD mosaics provided by the SSC were used.

\section{Spectral Fitting \label{sed_section}}
We fit the IRS and MIPS data, as well as additional photometry from
the NED, with models for the thermal and non-thermal emission, in
order to separate their respective contributions to the
mid-infrared. Simple functional forms are used to model the lobe and
jet synchrotron components, while the warm dust component is modeled
as an optically thin spherical shell of graphite and silicate grains
surrounding a source with an AGN accretion disk spectral energy
distribution. We also allow for the possibility that the dust
component is partially obscured behind a screen of cooler dust

For all of the objects in the sample with IRS spectra, we performed
fits to the SED using the following combinations of components:
\begin{itemize}\addtolength{\itemsep}{-0.7\baselineskip}
\item warm dust + lobe synchrotron;
\item warm dust + lobe synchrotron + jet synchrotron;
\item warm dust + lobe synchrotron + cool dust;
\item warm dust + cool dust + lobe synchrotron + jet synchrotron,
\end{itemize}
{\noindent with the free parameters varying simultaneously. In general, the combination of components which resulted in the best reduced chi-squared was selected as the best fit. Figure~\ref{examples} shows the best fits for the quasar 3C \,138 and the galaxy 3C \,340.

\section{Results\label{discussion}}

We detect powerful emission in the mid-infrared for our sample of
radio galaxies and quasars. All objects in our sample except one
were detected at 24\microns\ using MIPS, with luminosities
$L_{24\mu{\rm m}} > 10^{22.4}$\,W\,Hz\mo\,sr\mo . In the discussion
below, we characterize the infrared luminosities at rest-frame
wavelengths of 15 and 30\microns\ using the IRS and MIPS
measurements.

\subsection{Non-thermal Contribution\label{nontherm}}

Synchrotron emission from optically thin radio lobes has
flux-density spectral indices $\alpha \approx -0.7$, so the
contribution to the infrared is typically well over an order of
magnitude below the observed flux density. However, Doppler-boosted
synchrotron emission from dense, compact regions is a potentially
serious contaminant of the infrared emission. The relative
contribution of thermal and non-thermal processes to the infrared
emission of AGN has been the subject of many studies.

The {\em Spitzer\/} observations described here have provided
further constraints on the role of non-thermal processes in the
infrared emission of radio galaxies and quasars. In \S 4, we fitted the SEDs of the objects in our
sample with broadband spectral components representing synchrotron
emission from the radio lobes and jet as well as thermal emission
from circumnuclear dust (see Fig.~\ref{examples}). In this way, the
non-thermal contribution to the 15 and 30\microns\ luminosity was
estimated and subtracted from the observed emission. For the
galaxies, the thermal contribution is always $>80$\%; for the
quasars, the thermal contribution is in the range 10--100\%. All
objects, quasars and galaxies, were fitted in a consistent manner;
however, it is only the quasars in the sample which were estimated to
have a non-thermal contribution $>20$\%.

\subsection{Comparative Luminosity of 3C RR Radio Galaxies and Quasars\label{comparative}}

In order to compare the infrared luminosities of quasars and radio
galaxies we must first account for the variation of central engine
power amongst objects in our sample. To do this, we normalize the
infrared emission by the isotropic low-frequency radio luminosity to
get $R_{\rm dr} = \nu L_{\nu}(\rm IR)/ \nu L_{\nu}(\rm{178\,MHz})$.

The mean $R_{\rm dr}$ for quasars is $\sim 4$ times higher than
for galaxies at 15\microns . Subtracting the non-thermal emission
brings this down to $\sim 2$. The fitted optical depths allow us
to go further and ask whether obscuration by dust  can account for
this residual anisotropy of the thermal emission between quasars and
galaxies. Using the fitted screen optical depths to correct for dust
extinction, we find that the thermal emission from quasars and
galaxies is on average equal at 15\microns .

Figure~\ref{tauvsr} plots the screen optical depths, $\tau^{\rm
scr}_{9.7 \mu {\rm m} }$, against the core dominance parameter, $R =
F_{\rm core}/F_{\rm extended}$.  Our dust model (see
\S\,4) does not, a priori, impose any asymmetry in
the distribution of dust surrounding the nucleus.  From
Figure~\ref{tauvsr} there is evidence of an anti-correlation between
optical depth and core dominance, from which we can infer an
equatorial distribution of dust. The median screen (mixed) optical
depth for objects with ${\rm R} > 10^{-2}$ is 0.4 (1.0) and with
${\rm R} \leq 10^{-2}$ is 1.1 (3.0).

\section{Conclusions}

It is clear from previous work that beamed synchrotron emission and
dust extinction modulate the mid-infrared emission of quasars and
radio galaxies to some degree. In the work presented here, however,
we have quantified these effects for an orientation-unbiased sample
of powerful radio sources and shown that once they are taken into
account, quasars and radio galaxies are on average equally bright in
the mid-infrared.

\begin{figure}[!ht]
\centering

\plottwo{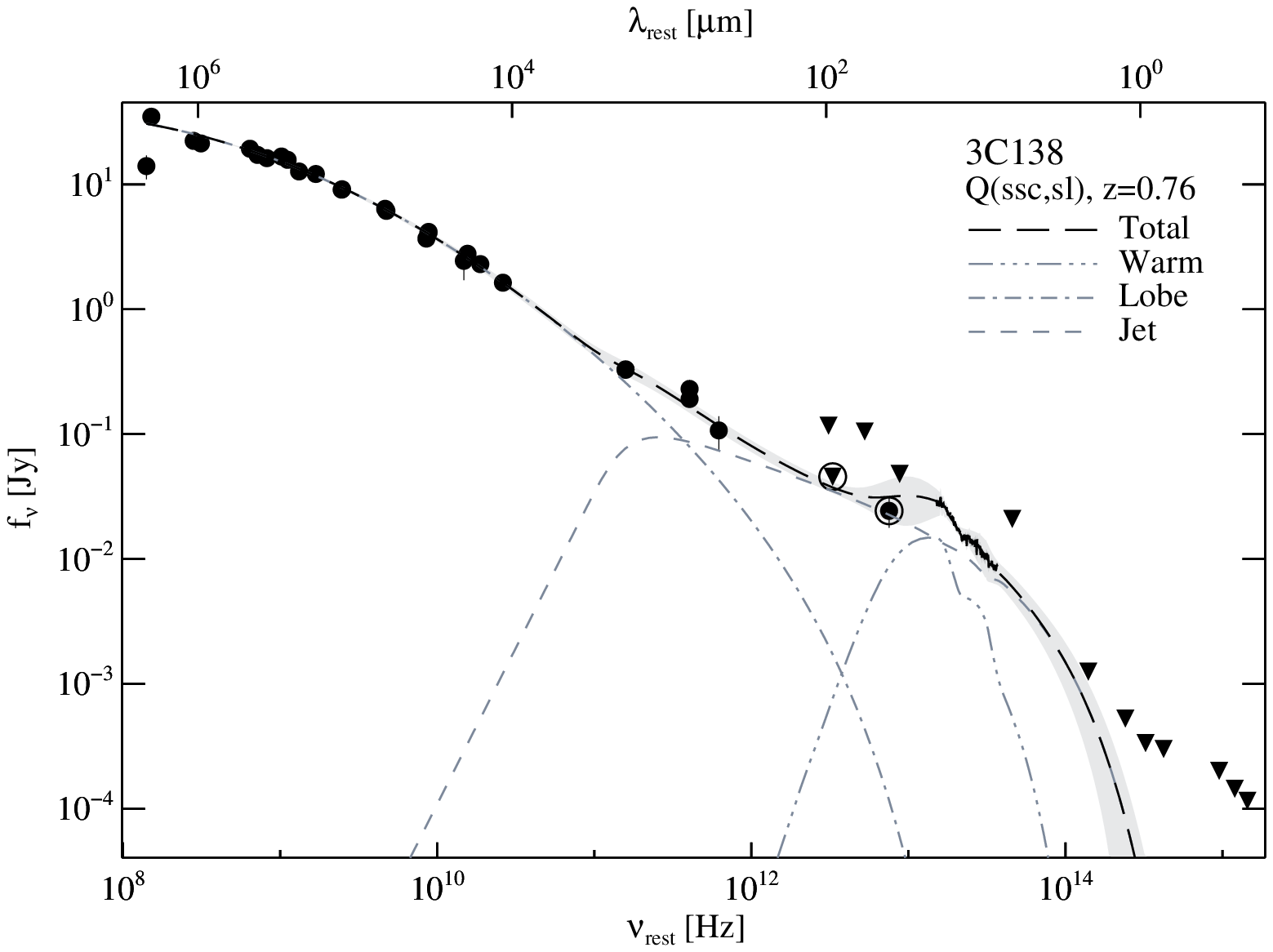}{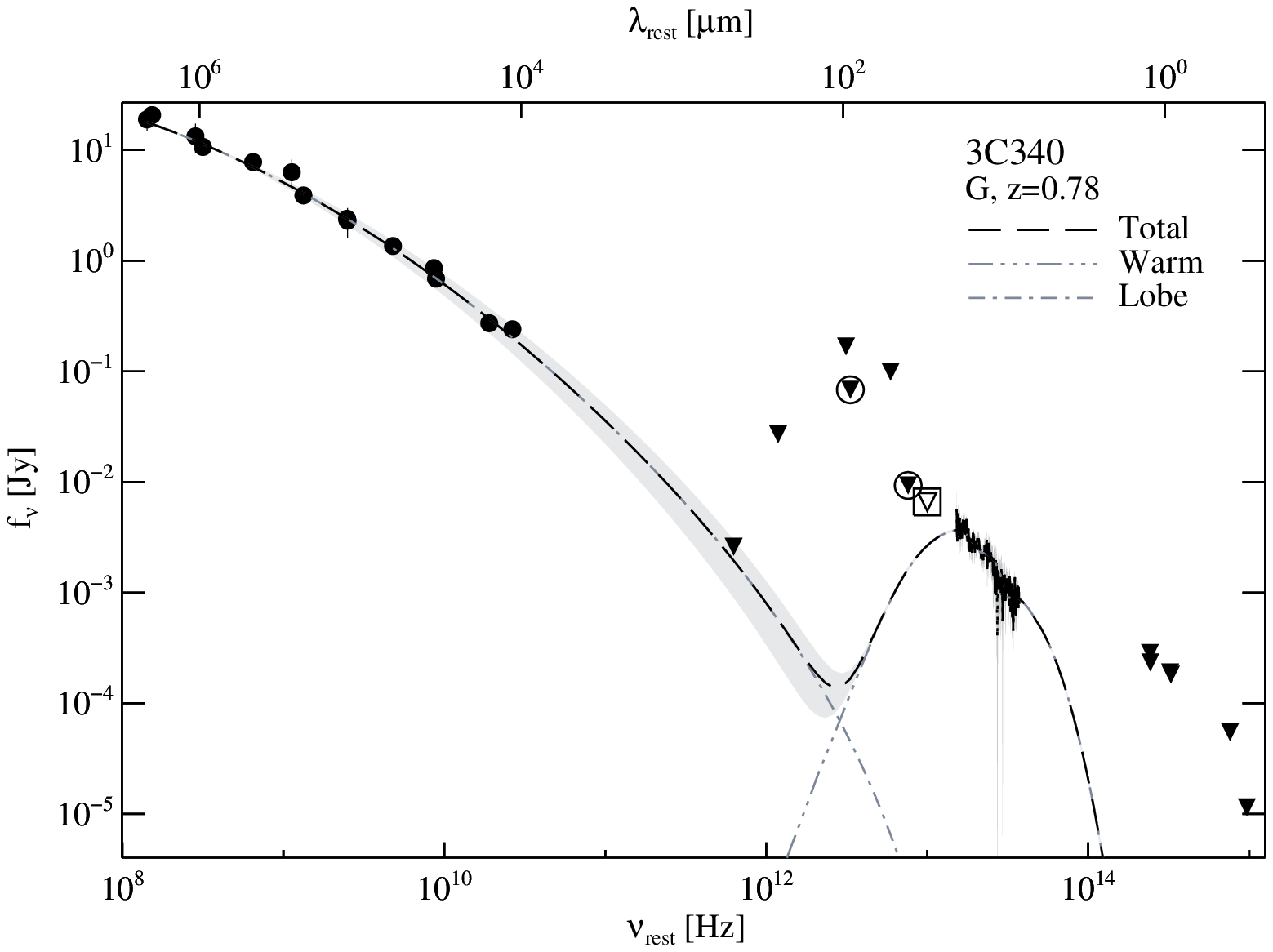}

\caption{The fits to the IRS spectra and photometric data for the quasar 3C\,138 (left) and galaxy 3C\,340 (right) are shown as examples. Filled circles and triangles represent detections and upper limits, respectively. Symbols within circles represent the MIPS measurements.
\label{examples}}
\end{figure}

\begin{figure}[!ht]
\centering
\includegraphics[scale=0.28]{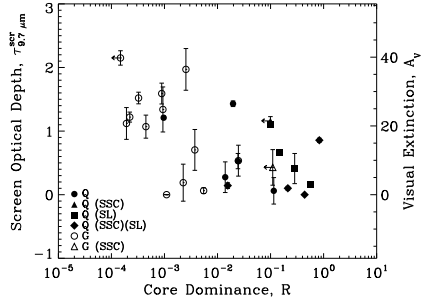}
\caption{The 9.7\microns\ screen  optical depths estimated from
the fits to the data versus the core dominance parameter, $R =
F_{\rm core}/F_{\rm extended}$. The right-hand $y$-axis indicates the visual extinction. From the apparent anti-correlation
of optical depth with core dominance, we infer an equatorial
distribution of cool dust consistent with the ``dusty torus'' of
orientation-based unification schemes.\label{tauvsr}}
\end{figure}

\acknowledgements 
The research described in this paper was carried out at the Jet Propulsion
Laboratory, California Institute of Technology, under a contract with the
National Aeronautics and Space Administration.

\end{document}